\documentstyle[twoside,fleqn,espcrc2,epsf]{article}
\newcommand{\be}{\begin{equation}}
\newcommand{\ee}{\end{equation}}
\newcommand{\bea}{\begin{eqnarray}}
\newcommand{\eea}{\end{eqnarray}}
\newcommand{\bean}{\begin{eqnarray*}}
\newcommand{\eean}{\end{eqnarray*}}

\newcommand{\AmS}{{\protect\the\textfont2
  A\kern-.1667em\lower.5ex\hbox{M}\kern-.125emS}}

\title{Exotic Quarkonia from Anisotropic Lattices}

\author{T. Manke\address{Center for Computational Physics,
University of Tsukuba, Tsukuba, Ibaraki 305, Japan}
for the CP-PACS Collaboration \thanks{A.~Ali~Khan, S.~Aoki, R.~Burkhalter,
S.~Ejiri, M.~Fukugita, S.~Hashimoto, N.~Ishizuka,  Y.~Iwasaki, K.~Kanaya,
T.~Kaneko, Y.~Kuramashi, T.~M., K.~Nagai, M.~Okawa, H.P.~Shanahan, A.~Ukawa 
and T.~Yoshi\'e}
}
       
\begin{document}

\begin{abstract}
We study in detail the spectrum of heavy quarkonia with different orbital angular
momentum along with their radial and gluonic excitations.
Using an anisotropic formulation of Lattice QCD we achieved an unprecedented
control over statistical errors and were able to study systematic errors such
as lattice spacing artefacts, finite volume effects and relativistic corrections.
First results on the spin structure in heavy hybrids are also presented.
\end{abstract}
%
\maketitle
\section{INTRODUCTION}
Heavy quarkonia are an interesting testing ground for our theoretical
understanding of QCD. Many models have been developed to describe
the vast amount of experimental data, but more accurate predictions
from first principles will be needed for future experiments with heavy quark
systems. However, the conventional lattice approach becomes overly expensive
if one  tries to accomodate all the non-relativistic energy scales on a single
grid: $Mv^2 \ll Mv \ll M$.
Here $v$ is the small velocity of the heavy quark with mass $M$.
The non-relativistic nature of such problems has frequently been employed
to formulate effective theories in which spatial and temporal directions
are treated differently at the level of the quark action \cite{NRQCD,FERMILAB}.
In particular the NRQCD approach has lead to very precise calculations
of the low lying spectrum in heavy quarkonia \cite{NRQCD_precision}

Additional problems arise if high energetic excitations are to be
resolved on isotropic lattices. Very often the temporal discretisation is
too coarse and the correlator of such heavy states
cannot be measured accurately for long times.
Glueballs and hybrid states are among the prominent examples of such
high-lying excitations and they receive much theoretical and experimental
attention as they are non-perturbative revelations of the gluon degrees of
freedom in QCD. Early lattice studies have predicted such states starting at
around 1.4 GeV, but those results were spoilt by large statistical
uncertainties \cite{early_glueball,cmhybrid,ukqcd_hybrid97,milc}.
This suggests that the inverse lattice spacing should at least be
3 GeV or higher. 
More recently anisotropic lattices have been used to circumvent this problem
in glueball calculations by giving the lattice a fine temporal resolution
whilst maintaining a coarse discretisation in the spatial direction 
\cite{morning_glue98}. The success of this approach has triggered new efforts
to measure hybrid potentials in an anisotropic gluon background and a
comparative NRQCD analysis has demonstrated the validity of
this adiabatic approximation for $b\bar bg$ hybrids. 
These attempts were reviewed in \cite{kuti_hybrid98}.

In a previous study we reported on first quantitative results 
for charmonium and bottomonium hybrid states from anisotropic lattices
\cite{cppacs_hybrid98}. Here we extend those methods to study
also other excitations and the spin structure in heavy quarkonia 
more carefully.

In Section 2 we present the details of our calculation and results for the
spin-averaged spectrum. In Section 3 we investigate the spin structure in
heavy quarkonia and report on novel results for the splittings in 
heavy hybrids and D-states.

\section{SPIN-AVERAGED SPECTRUM}
In order to study excited states with small statistical errors it is mandatory
to have a fine resolution in the temporal lattice direction, along
which we measure the multi-exponential decay of meson correlators.
To this end we employ an anisotropic and spatially coarse gluon action:

\begin{eqnarray}
S = - \beta \sum_{x, {\rm i > j}} \xi^{-1}\left\{\frac{5}{3}P_{\rm i j} -
\frac{1}{12}\left(R_{\rm i j} + R_{\rm j i}\right)\right\} \nonumber \\
- \beta \sum_{x,{\rm i}} \xi \left\{\frac{4}{3}P_{\rm i t} -
\frac{1}{12}R_{\rm i t}\right\}~~.
\label{eq:glueaction}
\end{eqnarray}
Here $(\beta,\xi)$ are two parameters, which determine the gauge coupling
and the anisotropy of the lattice. Action (\ref{eq:glueaction}) 
is Symanzik-improved and involves plaquette terms, $P_{\mu\nu}$, as well as
rectangles, $R_{\mu\nu}$. It is designed to be accurate up to ${\cal
O}(a_s^4,a_t^2)$, classically. 
To reduce the radiative corrections we invoked mean-field improvement
and divided all spatial and temporal links by a 'tadpole' coefficient
$u_{0s}$ and $u_{0t}$, respectively.
We determined those coefficents self-consistently by measuring spatial 
and temporal plaquettes: $u_{0s}= \langle {\rm Tr}~P_{ij}\rangle ^{1/4} $ and
$u_{0t}= \langle {\rm Tr}~P_{it}\rangle ^{1/4}$.
With this prescription
we expect only small deviations of $\xi$ from its tree-level value $a_s/a_t$.

To describe the forward propagation of heavy quarks 
in the gluon background ($A_\mu$) we used the NRQCD approach introduced in
\cite{NRQCD}: 
\be
G_{t+a_t} = \exp{\bigg[-a_t~\left(H(g,M)+igA_t\right)\bigg]}~G_{t}~~.
\label{eq:prop}
\ee
Here the NRQCD Hamiltonian, $H$, is designed to account for relativistic
corrections and includes spin-dependent operators up to ${\cal O}(mv^6)$.
The implementation details of Equation (\ref{eq:prop}) on anisotropic lattices 
are given in \cite{ron_lat98}.
Since we are working with spatially 
coarse lattices it is crucial to improve all lattices derivatives and
colour-electromagnetic fields in Equation (\ref{eq:prop}). 
Following the prescription of \cite{NRQCD} we also achieved
an accuracy of ${\cal O}(a_s^4,a_t^2)$ in the quark sector.

At each value of the coupling, $\beta$, we carefully tuned the heavy 
quark mass, $M$, so as to reproduce the experimental ratios $M_{kin}/(1P-1S)$
very accurately. Fortunately, the spin-independent quantities, such as $1P-1S$,
are not very sensitive to the actual value of $M$, but the spin structure 
will show a strong dependence.
For our calculation we chose several different values of the coupling, 
$\beta$, which correspond to spatial lattice spacings between 0.15 fm and 0.47
fm. On even coarser lattices we cannot expect to control discretisation
errors with our simple-minded approach, while much finer lattices will violate the
validity of NRQCD which requires $a_sM > 1$.

From the quark propagator, $G_t$, we construct meson correlators for bound
states with spin $S=(0,1)$ and orbital angular momentum $L=(0,1,2)$.
Magnetic hybrid states are constructed from the colour-magnetic
field coupled to a $Q\bar Q$-pair ($B_i = [\Delta_j,\Delta_k]$).
For example, the spin-singlet operators read
\be
\bar Q^\dag Q~,~
\bar Q^\dag \Delta_i Q~,~ 
\bar Q^\dag \Delta_j \Delta_k Q \mbox{~~and~~}
\bar Q^\dag B_i Q~.  
\ee
Those simple operators can be further improved upon in order to optimise
their overlap with the states of interest. Here we extract the excitation
energies from multi-exponential fits to several different correlators.
 
\begin{figure}[htb]
\hbox{\epsfxsize = 80mm \epsfysize = 60mm \hskip -5mm \epsffile{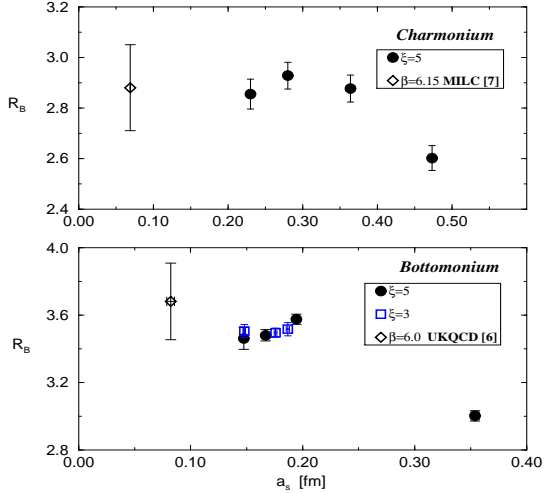}}
\vskip -5pt
\caption{Scaling analysis for spin-averaged hybrids. 
As we only measure excitation energies relative to the
ground state it is natural to present our results as the ratio
$R_B=(1B-1S)/(1P-1S)$, which gives the normalized splitting of the magnetic 
hybrid excitation above the 1S.}
\label{fig:RB}
\vskip -20pt
\end{figure}

Within the NRQCD approach one cannot extrapolate to the continuum limit 
and it is paramount to establish a scaling region
for physical quantities already at finite lattice spacing.
In a previous study we found such scaling windows
for the spin-averaged gluon excitations in both charmonium and bottomonium
\cite{cppacs_hybrid98}.  
These results are particularly encouraging as
they are in excellent agreement with calculations on isotropic lattices 
\cite{ukqcd_hybrid97,milc}, but with much smaller errors.
In addition we were able to check our predictions against possible systematic
errors such as finite volume effects. This is a natural concern since hybrids
are expected to be rather large owing to the flat potentials they are living
in. In Figure \ref{fig:RB} we show our results from lattices all larger 
than 1.2 fm in extent, beyond which we could not resolve any volume dependence.
For the Bottomonium hybrid we also have consistent results
from two different anisotropies ($\xi=3,5$) which confirms our initial assumption of 
small temporal lattice spacing artefacts. In fact, on our lattices
the tadpole coefficient $u_{0t}$ deviates from its continuum value 
by only about 5\% or less. 
It is also interesting to notice the presence of scaling violations 
which are clearly visible if the lattices are too
coarse for the physical system.

\begin{figure}[htb]
\hbox{\epsfxsize = 80mm  \epsfysize = 60mm \hskip -5mm \epsffile{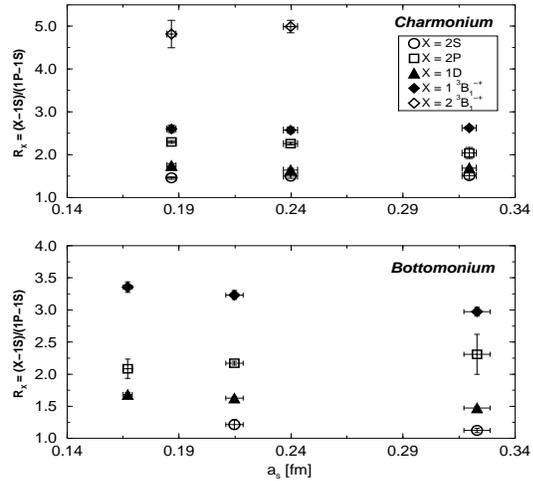}}
\vskip -5pt
\caption{Scaling analysis for excited Quarkonia. We plot the ratio $R_X$
against the spatial lattice spacing for different states X = S,P,D
and ${^3}B_1^{-+}$ (exotic hybrid).}
\label{fig:RX}
\vskip -20pt
\end{figure}

We have now extended our analysis to study also higher radial excitations and
D-states with $L=2$. The spin-independent results are summarised in Figure
\ref{fig:RX}.
The possibility to resolve all these excitations reliably should be 
considered the main success of anisotropic lattices. 
We are presently performing a finite volume analysis of the higher radial
excitations. 
However, with the newly achieved accuracy we can also study
spin-splittings in more detail.

\section{SPIN STRUCTURE}
Our inclusion of relativistic corrections to Equation (\ref{eq:prop}) is a
significant improvement over previous NRQCD calculations of hybrid states,
which were restricted to only leading order in the velocity expansion: 
${\cal O}(mv^2)$.  At this level there are no spin-dependent operators
and we have a strict degeneracy of all singlet and triplet states.
Within the NRQCD framework, the spin-dependent operators appear first as 
higher order corrections to the Hamiltonan: ${\cal O}(mv^4)$. 
This is in accordance with the experimental observation that spin-splittings
in quarkonia are suppressed by $v^2$ compared to the spin-independent 
structure discussed in the previous section. 
Here we also include spin correction terms up to ${\cal O}(mv^6)$,
and study the breaking of the degeneracy.
In particular, we could directly observe the exotic hybrid, $1^{-+}$,  
which is the state of greatest phenomenological interest.

\begin{figure}[t]
\hbox{\epsfxsize = 80mm  \epsfysize = 60mm \hskip -1mm \epsffile{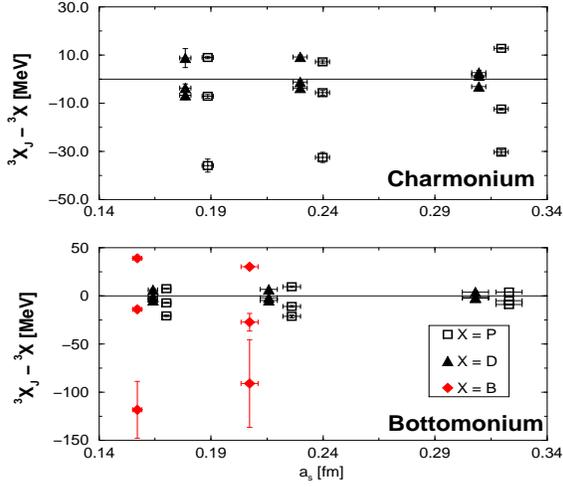}}
\vskip -5pt
\caption{Fine Structure in Charmonium and Bottomonium on three different
lattices. All results come from NRQCD calculations with accuracy ${\cal
O}(mv^6,a_s^4,a_t^2)$. The conversion to dimensionful units is done by setting
the $1P-1S$ splitting to its experimental value.} 
\label{fig:fs}
\vskip -20pt
\end{figure}

\begin{figure}[t]
\hbox{\epsfxsize = 80mm  \epsfysize = 60mm \hskip -3mm \epsffile{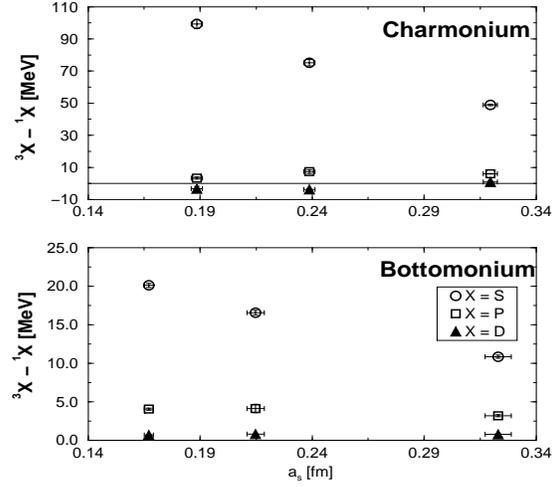}}
\vskip -5pt
\caption{Hyperfine splitting between the averaged spin-triplet and
spin-singlet states of different orbital angular momentum.}  
\label{fig:hfs}
\vskip -20pt
\end{figure}

Our results for fine structure and hyperfine splittings 
are shown in Figure \ref{fig:fs} and \ref{fig:hfs}, respectively.
There are several interesting observations one can make.
First of all, we find a noticeable reduction of the fine structure
in D-states when compared to that of P-states. Similarly, the hyperfine
splitting between spin-singlet and spin-triplet states is
equally suppressed as the orbital angular momentum is increased. 
This is in accordance with potential models which predict a 
hyperfine splitting for only (L=0)-states since all other wavefunctions 
vanish at the origin.
Our data also indicates that the fine structure in hybrid states 
is enlarged compared to the splittings in P-states, but we could not yet
resolve any splitting between the spin-triplet state (${^3}B = 5~{^3}B_2 +
3~{^3}B_1 + {^3}B_0$) and the singlet ${^1}B_1$.

Finally one should also notice that the scaling behaviour
of the spin structure is more involved than that of the spin-independent
spectrum. This is not surprising since we have adopted a very simplicistic
approach to determine all the coefficients in NRQCD with a single prescription
(tree-level tadpole improvment).  Namely the hyperfine splitting
${^3}S_1-{^1}S_0$ does not scale on the lattices considered
here. It is apparent that one needs a better improvement description to
account for lattice spacing artefacts in such UV-sensitive quantities.

In conclusion, we find that coarse and anisotropic lattices
are extremely useful for precision measurements of higher excited states.
This is due to an improved resolution in the temporal direction
and the possibility to generate large ensembles of gauge field configurations 
at small computational cost.
It has lead to an unprecedented control over statistical and systematic
errors in lattice studies of heavy quarkonia. 
After the inclusion of relativistic corrections we are also sensitive 
to spin-spin and spin-orbit interactions and could observe a clear hierachy in
the spin structure, depending on the orbital angular momentum. 
Spin splittings in hybrid states were found to be larger as a result of the
gluon angular momentum to which the spin can couple. The remaining and
dominating systematic error for all our predicitions is an uncertainty in the scale as the result of the quenched approximation. 
This is not yet controlled and we find a variation of 10-20\%,
depending on which experimental quantity is used to set the scale.

This work is supported by the Research for the Future Programme of the JSPS.


\begin{thebibliography}{9}
\bibitem{NRQCD} G.P. Lepage {\it et al.} Phys.Rev.{\bf D}46 (1992) 4052. 
\bibitem{FERMILAB} A.X. El-Khadra {\it et al.},  Phys.Rev.{\bf D}55 (1997) 3933. 
\bibitem{NRQCD_precision} C.T.H. Davies {\it et al.}, Phys.Rev.{\bf D}50
(1994) 6963; T. Manke {\it et al.}, Phys.Lett.{\bf B}408 (1997) 308.
\bibitem{early_glueball} J. Sexton {\it et al.}, Phys.Rev.Lett.75 (1995) 4563.
\bibitem{cmhybrid} S. Perantonis and C. Michael, Nucl. Phys. {\bf B}347 (1990) 854.
\bibitem{ukqcd_hybrid97}T. Manke {\it et al.}, Phys.Rev.{\bf D}57 (1998) 3829.
\bibitem{milc} C. Bernard {\it et al.}, Phys.Rev.{\bf D}56 (1997) 7039.
\bibitem{morning_glue98} C.J. Morningstar and M. Peardon, Phys.Rev. {\bf D}56
 (1997) 4043.
\bibitem{kuti_hybrid98} J. Kuti, Nucl.Phys.(Proc.Suppl)73 (1999) 72.
\bibitem{cppacs_hybrid98} T. Manke {\it et al.} (CP-PACS Collaboration), Phys.Rev.Lett.82 (1999) 4396.
\bibitem{ron_lat98} I.T. Drummond {\it et al.}, Nucl.Phys.(Proc. Suppl.) 73 (1999) 336.
\end{thebibliography}
\end{document}